# *Limitless FaaS: Overcoming serverless functions execution time limits with invoke driven architecture and memory checkpoints*

Rodrigo Landa Andraca
Escuela de Ingeniería y Ciencias
Tecnológico de Monterrey
Jalisco, México
rodrigo.landa@tec.mx

Mahdi Zareei
Escuela de Ingeniería y Ciencias
Tecnológico de Monterrey
Jalisco, México
mahdi.zareei@tec.mx

*Abstract*—Function-as-a-Service (FaaS) allows to directly submit function code to a cloud provider without the burden of managing infrastructure resources. Each cloud provider establishes execution time limits to their FaaS offerings, which impose the risk of spending computation time without achieving partial results. In this work, a framework that enables limitless execution time in FaaS, with little to no modifications to the user-provided function code, is presented. After a thorough literature and theoretical framework review, Apache OpenWhisk Actions and the DMCTP checkpoint-and-restore (CR) tool were selected. With these, dependent successive serverless same-function invocations that exploit the persistence of partial results were implemented. The solution was submitted to the FaaSDom benchmark and time metrics were collected. Additionally, the solution was characterized in terms of the Serverless Trilemma. The resultant system, even at this proof-of-concept state, offers a lot of value to companies that rely heavily on serverless architecture.

## I. Introduction

Serverless computing allows efficient development and deployment without the burden of managing infrastructure resources. The provider oversees automatic scaling and bills only for execution time. Function-as-a-Service (FaaS) is but one of the many forms serverless takes, and it allows to directly submit function code to a cloud provider.

Each cloud provider imposes CPU, memory, and execution time limits to their FaaS platforms. This forces developers to re-architect or extend their software, having as consequence an additional overhead not found in serverful implementations that can represent a considerable expense for technology companies.

Reducing the chance of spending computing resources without at least getting partial results may bring important savings. This not only translates into smaller cloud budgets, but also allows organizations to generate more job openings and offer their employees more competitive salaries. Solving this problem brings benefits to cloud providers too, since cloud architects and developers may walk away from FaaS due to its restrictions. Additionally, more effective computing time implies a reduced carbon footprint.

In this work we present an approach to FaaS that enables limitless execution time when processing workloads, with little to no modifications to the user-provided function code. For its validation, we measured the execution time of the solution, characterized it in terms of Serverless Trilemma [2] violations, and documented the user experience affections.

Section II goes through a background on FaaS limitations and works seeking to prevent them or diminish their impact, and it introduces the main technologies required for the system's implementation. The containers, OpenWhisk, and DMTCP topics from this section are particularly important. The implementation of the solution, the experimental setups, and benchmarks scripts are described in section III. On section IV, results from the benchmarks are discussed and the solution is characterized in terms of Serverless Trilemma [2] violations. Critical factors that affect the final user experience are also mentioned in this chapter. Finally, on section V conclusions and future work are presented.

## II. Background and Related Work

### A. Background

*a) Containers and Docker:* Containers are similar to Virtual Machines but don't require their own operating system because they share the host's kernel. This allows containers to free up resources like storage, CPU, and RAM, keep a small size, and startup quickly. As a consequence, containers are perfect for portability; software code can be packaged together with the related configuration files, libraries, and other runtime dependencies. The era of containerization ushered thanks to the Docker Engine, an open-source software to create, manage, and orchestrate containers. It works on the principle of a client-server application, where the server runs the actual Docker daemon process and the client is the Docker command-line interface. Containers are generated from images, layers of files and directories that provide a read- only composite filesystem as a starting point. These layers are stacked one upon the other in a particular order, so that each stores only the differences from the layer below it. When creating a container from an image, a thin read-write layer is added on top that enables copy-on-write semantics. Images are generated by committing a container or from a Dockerfile (text file with commands to assemble an image). When building an image, there are multiple optimization techniques, like build cache exploitation and multi-stage builds. Once built, images are stored in public or private registries that can be hosted by a third party. The most popular by far is DockerHub [13].

*b) OpenWhisk:* OpenWhisk is a serverless, open source cloud platform. It follows an event-driven architecture, running Actions according to Rules that apply to Events channeled through Triggers. Internally, OpenWhisk exploits technologies like nginx, Kafka, Docker, and CouchDB. These components interact as follows: 1) the nginx server receives a request to trigger an Action, 2) the request is forwarded to a Scala REST API named Controller, 3) authentication and authorization are performed considering the subjects database in a CouchDB instance, 4) the requested action is loaded from the whisks database, 5) a Scala load balancer choses one of the available Invokers, 6) a Kafka message addressed to an Invoker is published, 7) the Invoker (also developed in Scala) injects the action code in a Docker container and executes it, and 8) the results are stored in the activations database [14]. Actions can come from a function programmed in one of the supported runtimes, a compatible binary executable, or from Docker containers packaged with an executable. The latter must implement the Action Interface by exposing an HTTP server on port 80 with `/init` and `/run` endpoints that can handle invocation requests. To make this easier, a DockerHub-available Apache OpenWhisk public image can be used as a starting point [15]. As any other cloud platform, OpenWhisk imposes certain system limits. For example, Actions' timeout has a default value of 60000 ms (60 s), but it can be pushed up to 300000 ms (300 s). When such a limit is reached, the container is terminated. Likewise, Actions' memory can be of 512 MB at most, being 256 MB when not specified [14].

*c) IBM Cloud:* IBM Cloud is a set of cloud computing services offered by IBM. These include Cloud Functions, Cloud Object Storage, and Cloudant. Cloud Functions is a FaaS programming platform based on OpenWhisk. It supports the most popular programming languages and runs code in response to HTTP API requests, IBM Cloud services events, and third-party events [16]. Cloud Object Storage is a highly available, durable, and secure platform for storing unstructured data. Files (called objects here) are organized into buckets, where the hierarchy is effectively flat [17]. Cloudant is a JSON document store available as a service. It is based on Apache CouchB, but doesn't require any installation, server management, or configuration setting [18].

*d) DMTCP:* Distributed MultiThreaded CheckPointing (DMTCP) is a transparent user-level (no system privileges required) checkpointing open-source package for distributed applications. It is designed to support high performance applications, typical desktop applications, and long-running distributed applications. To use it, a command line interface is provided. Its most important options are `dmtcp_checkpoint` — to register a process as one of the set of child processes that will be checkpointed — and `dmtcp_comand` — to write checkpoint images for each registered process. The `dmtcp_restart_script.sh` bash script generated at checkpoint time is also important, since it contains all the commands needed to restart the computation. Its software architecture is based on two layers: MTCP (for single process checkpointing) and DMCTP itself (to checkpoint a network of processes spread over many nodes). These layers' entrypoint is a shared library written in C and C++ that gets injected to arbitrary applications at execution time. This library loads MTCP to create the checkpoint manager thread and enables the integration with DMCTP itself. At this time, DMTCP opens a TCP/IP connection to the checkpoint coordinator and adds wrappers around libc functions to be aware of all forked child processes. The checkpointing distributed algorithm is executed asynchronously in each user process and only uses a cluster-wide barrier as communication primitive [19].

*B. Related Work*

*a) FaaS constraints and workarounds:* Kuhlenkamp et al. [1] identified disadvantages current FaaS offerings have. One of them is the time limit imposed by cloud providers, which computationally intensive applications can easily exceed. This boundary may be mitigated, they claim, by chaining multiple executions of the same function handler for the same event. Such an approach, like others introduced later, violates at least one of the constraints of the Serverless Trilemma (ST). The trilemma states that when using FaaS to implement function compositions only two of three constraints—double billing, black box, and substitution—can be satisfied [2]. *Sequential Workflow in Production Serverless FaaS Orchestration Platform* took the ST in consideration when comparing the execution time of the reflection, fusion, chaining, async, and client sequential composition patterns. As expected, one or more of the ST constraints are violated by these approaches [3]. Taibi et. al [4] identified and classified patterns for FaaS. In the orchestration and aggregation category they included the fan-in/fan-out and function chain patterns to enable the execution of long tasks that exceed the maximum execution time. However, these entail strong coupling and can't avoid the complexity of splitting tasks. An interesting observation this work makes is that many patterns have been created exclusively to work around serverless limitations. Other patterns to overcome FaaS limitations are described in *A shared memory approach for function chaining in serverless platforms* and *Resource-Centric Serverless Computing*. In the former, shared-memory direct function-to-function communication implemented in Docker containers is proposed for function chaining [5]. In the latter, a serverless computing platform that models and executes applications in a resource-decoupled way is presented. With their OpenWhisk-based ReSC platform, an application is modeled as a resource graph with components of arbitrary size and duration [6].

*b) Benchmarks and testbeds:* The aforementioned patterns to overcome FaaS limitations need to be evaluated in a standard manner, that's why there are many documented academic efforts to generate benchmark suites. Copik et al. [7] designed SeBS, a benchmark suite composed of a collection of serverless applications oriented to different performance profiles, for which local (time, CPU utilization, memory, I/O, and code size) and cloud (microbenchmarking, provider and client time, memory, and cost) metrics are collected. The serverlessbench benchmark suite focuses on metrics that are unique and significant to serverless platforms. From Yu et al.'s [8] reported executions, a series of implications that can aid the design of serverless systems were derived: decoupling a serverless application with varied resource needs across execution phases might save costs; nested function chains require more resources and execution time, and bare a high timeout risk; composition methods can significantly impact the billing in serverless computing; functions with larger code sizes should be optimized to avoid longer startup latencies; and improved execution performance can be achieved in serverless platforms by sharing the implicit states among instances of a function. Finally, the FaaSdom benchmark suite consists of a collection of tests targeting CPU, network latency, and disk IO performance [9]. After running the tests, its authors concluded that FaaS is heavily limited by the available runtime systems and programming languages, supported triggers, third party services integrations, and quotas or maximum memory.

*c) Checkpointing:* As stated in [10], checkpointing can be performed on system or application level. On system level, no code changes are required, full program states are saved, and after a failure the program must be restarted from the last checkpoint. In contrast, on application level only user-defined data is checkpointed, and requires some programming effort. Both of these contribute towards fault tolerance. For example, on [11] checkpointing of a network of virtual machines is used to provide fault tolerance to complex distributed applications. The approach introduced allows such a network to be started locally, checkpointed, and later re-deployed or resumed in a cloud platform. The implementation is based on DMTCP. A. Ahmed et al [12] also used DMTCP for checkpointing, but in a fog computing Docker-based system. The significant amount of time required to boot a Docker container must be taken into account in a fog computing environment, where a given container may be repeatedly launched, created, and booted. This work reduces the boot time impact by restarting containers from fully-booted checkpointed states. The system consists of two components: 1) a thin container to checkpoint and restart an application with its environment, and 2) a mechanism that leverages Ceph distributed storage to share the container environments and checkpoint images.

## III. SYSTEM DESIGN

The FaaS OpenWhisk offerings were used for the solution. On it, the DMCTP checkpoint-and-restore (CR) tool was tested to guarantee its compatibility with the execution environment. Finally, dependent successive serverless same-function invocations that exploit the persistence of partial results were implemented.

Figure 1 provides an overview of the solution's architecture. As it can be seen, we have a workload that requires a computation time greater than a single function timeout. $f_1$ is the first instance of the serverless function with the code to process the workload. It uses the CR tool to generate a checkpoint before timing out and invokes $f_2$, another instance of the same serverless function. $f_2$ restores $f_1$'s endpoint and then contributes to the workload processing progress before checkpointing and invoking $f_3$. These subsequent invocations of the same function that perform checkpoint and restore operations continue until the whole workload is computed.

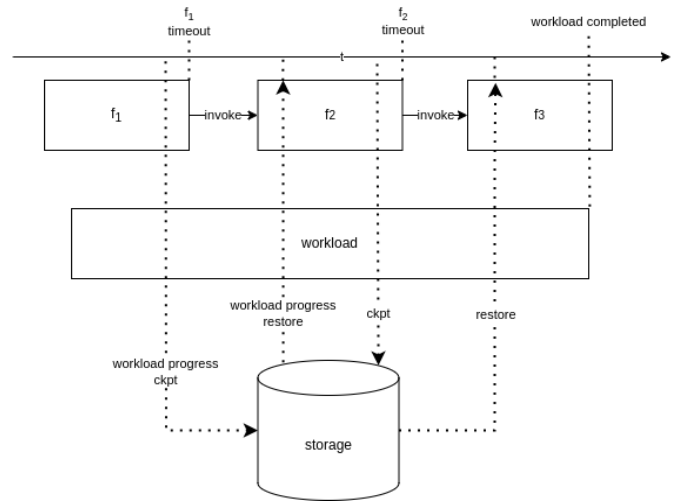

Fig. 1. Solution architecture.

### A. System implementation

In the following sections, the OpenWhisk Action's implementation details are introduced.

*a) DMTCP and Python Docker image:* According to DMTCP installation instructions, a Debian-based operating system and the git, gcc, g++, and make dependencies are required [20]. Additionally, Python 3.9 is installed for the Action's script that handles OpenWhisk requests. The resultant Dockerfile is occupied to generate the `dmtcp_python` image, on which the next section builds upon.

*b) OpenWhisk Action custom runtime image:* The actual custom runtime image for the OpenWhisk Action is generated from the `dmtcp_python` image. This image also includes multiple Python packages. Some are required by the Flask server that handles the OpenWhisk requests, and some are used to store the checkpoint files remotely. The source code for the Flask server, the OpenWhisk Action's runner implementation, the checkpoint files repository, the sequential invoker service, the results repository, and the target functions to be submitted to checkpointing are all added to the image too. Furthermore, the DMTCP checkpoint directory, logging level and directory, and the Flask server port are set via environment variables. OpenWhisk parameters for sequential action execution are also configured with environment variables. As the container's entrypoint, a script that simply starts the Flask server after exporting the DMTCP_COORD_HOST environment variable is used. This needs to occur at the entrypoint because each container will likely be allocated in a different host.

   *c) Action Runner:* The Action Runner is implemented as a Python class with run and init methods that get called when the respective /init and /run Flask paths receive requests. The skeleton for Docker Actions [20] had to be modified to exploit DMTCP. There are three additions worth mentioning, all of them pertaining to the run method. The first one consists of a function for persisting the checkpoint and invoking the same action, which gets called after some time that should be configured to be smaller than the Action's timeout (50 seconds in this case). Such scheduled execution must be canceled if the Action completes or errors out. The second one is required to download the checkpoint from the repository and restart it, if it is available. The last one is for launching the binary wrapped by DMTCP.

### B. Experimental setup

   *a) Implementation specifics:* The solution can be executed wherever an OpenWhisk platform can be set up. In this case, a local standalone OpenWhisk stack and IBM Cloud Functions are considered. For each of them, different implementations of the checkpoint repository, invoker service, and results repository are occupied. Table I summarizes the implementation details. Given that each implementation needs particular parameters, distinct images with specific arguments are used for each setup. For the deployment of infrastructure on IBM Cloud, terraform configuration files for the Function Action, the Cloud Object Storage bucket, and the Cloudant database are used.

TABLE I. REPOSITORIES AND SERVICES IMPLEMENTATION DETAILS PER EXECUTION ENVIRONMENT

| Setup | Implementations | | |
|---|---|---|---|
| | *Checkpoint repository* | *Invoker Service* | *Results repository* |
| Local standalone stack | Local SFTP server mounted in the same host | REST API request to OpenWhisk API in the same host | Local SFTP server mounted in the same host |
| IBM Cloud Function | IBM Cloud Object Storage | REST API request to IBM Cloud Function's OpenWhisk API | IBM Cloudant |

   *b) Test target function:* The same test target function is used in both local and cloud environments by default. This function counts to 70, waiting 1 second between each increment. In every multiple of 10, a HTTP POST request to an external server's /count resource is performed. When the counting ends, a final HTTP POST request is made to the /result resource.

### C. Benchmarks

To benchmark the solution, the FaaSdom suite provides tests targeting CPU and memory performance. Its faas-fact memory-bound and faas-matrix-mult CPU-bound tests respectively factorize an integer and multiply large integer matrices [9]. In the suite's repository, Python implementations for IBM Cloud Functions are provided [21].

To execute a specific function, its name and parameters must be passed in the invoke request. Running matrix requires, for example, the following wsk command line interface call:

```
wsk action invoke wsk_dmtcp_python
--result --param bin matrix
--param bin_args [1]
```

### IV. RESULTS AND DISCUSSION

#### A. Benchmark results

Benchmarks were executed using a script. Through it, the factors and matrix binaries were executed 20 times with different arguments. In each run, the execution time and the number of function invocations were captured. It is worth mentioning that the Action timeout was set to 90 seconds.

The tests were performed on a local OpenWhisk standalone stack, running on an Intel® CoreTM i7-8565U CPU @ 1.80GHz × 8, 16.0 GiB, Ubuntu 22.04.3 LTS laptop. Figures 2 and 3 summarize the results obtained.

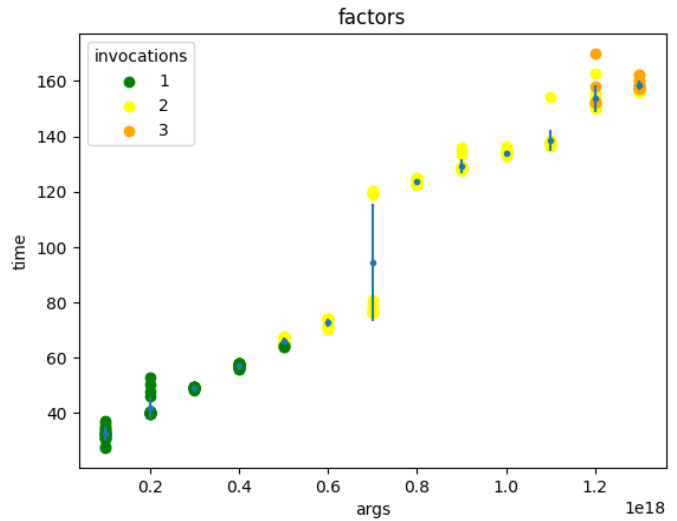

Fig. 2. factors execution time and number of invocations for multiple runs of various arguments.

In Figure 2, it can be observed that some runs of the factorial of $0.5 \times 10^{18}$ required 2 lambda invocations. Likewise, when calculating for $1.2 \times 10^{18}$ some required 2 invocations, and some required 3. Similarly, (as Figure 3 demonstrates) when multiplying $900 \times 900$ matrices a fraction of the runs required 2 lambda invocations, and $1100 \times 1100$ operations sometimes took 3 invocations and sometimes even 4.

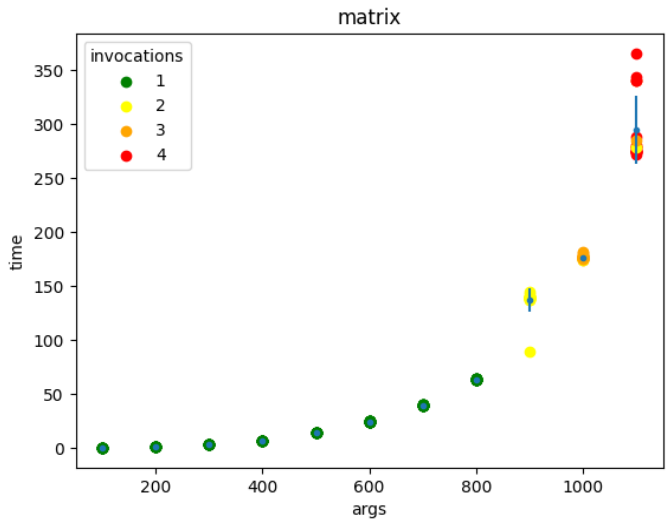

Fig. 3. matrix execution time and number of invocations for multiple runs of various arguments.

Having the execution time increase as arguments do is expected. More importantly, in all executions the service managed to checkpoint the system and restore it in a sequentially invoked function. However, there were some cases in which the executable managed to finish while the checkpoint was being made. This provoked a sequential function invocation and generated two results with different execution times.

Figures 4, 5, 6, and 7 summarize the mean CPU and RAM usage percentage in the last invocation as the `factors` and `matrix` arguments vary.

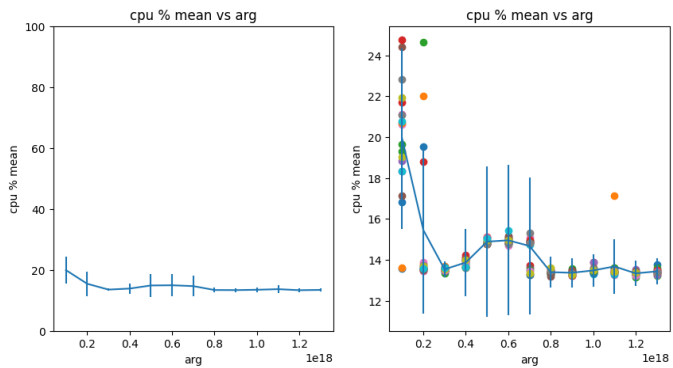

Fig. 4. Mean CPU usage percentage in factors' last invocation as the argument increases.

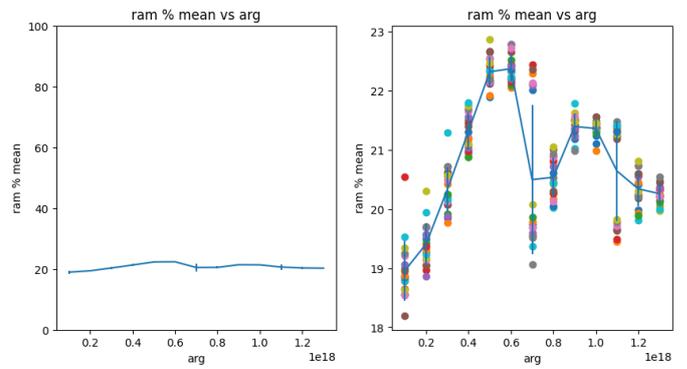

Fig. 5. Mean RAM usage percentage in factors' last invocation as the argument increases.

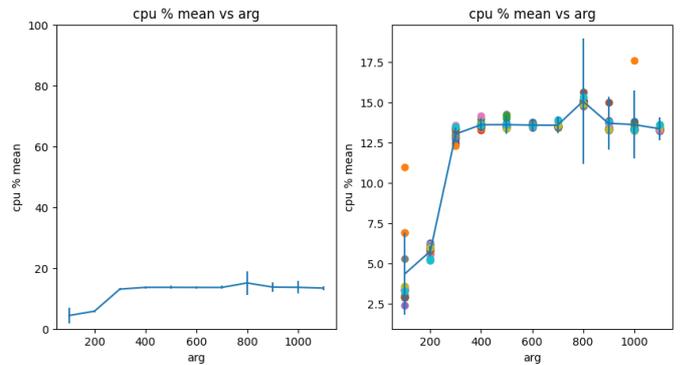

Fig. 6. Mean CPU usage percentage in matrix's last invocation as the argument increases.

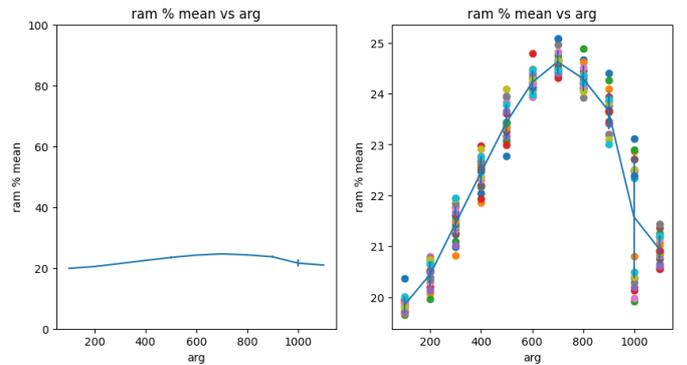

Fig. 7. Mean RAM usage percentage in matrix's last invocation as the argument increases.

The `factors` overall mean CPU usage percentage and RAM usage percentage were 14.46% and 20.89% respectively. On the other hand, the `matrix` binary had a 13.87% overall mean CPU usage percentage and a 23.22% overall mean RAM usage percentage. It is worth noting that there was no evident CPU or RAM usage percentage increase as the arguments (and consequently number of invocations) did. This can be interpreted as no affection to CPU or RAM consumption as the number of invocations adds up.

The benchmark scripts didn't require any alteration from the FaaSdom repository. Although the same can't be said for the other two benchmarks in the repository. `faas-netlatency` requires additional network libraries and idempotence in the remote service being called to guarantee correctness. On the other hand, `faas-diskio` would need to generate files of at least hundreds of MBs to achieve runs longer than 60 seconds. Files of such size demand a considerable amount of network bandwidth for transference and add to checkpointing time. The disk paths that each cloud platform makes available for temporary file storage would need to be parameterized too.

*B. Serverless Trilemma violations*

As previously mentioned, the Serverless Trilemma states that when using FaaS to implement function compositions only two of three constraints — double billing, black box, and substitution — can be satisfied.

This solution violates double billing, because the function may be invoked twice and even thrice (see Figures 2 and 3). In some cases, it violates the black box constraint too, such as in the filesystem and network latency tests, and any other executable that requires specific code changes. Nevertheless, it does comply with the substitution principle, since the function can still be used for binaries that require a single execution.

*C. Final user experience*

The following points affect the final user or developer experience:

- The executable's programming language is not limited to Python, but it must be supported by DMCTP.

- Logging proved to be complicated, as logs from the subprocess orchestrated by DMTCP must be captured and forwarded to the Action Runner's STDOUT.

- Checkpointing time must be taken into consideration. The number of files to checkpoint and the network latency to reach the central storage are two important factors.

- Function code must be packaged as an image, as there is no other way to include DMTCP.

V. CONCLUSIONS AND FUTURE WORK

*A. Conclusions*

The objective of implementing a tool for FaaS that enables limitless execution time when processing workloads, with little to no modifications to the user-provided function code, was achieved. Furthermore, its execution time was benchmarked, it was characterized in terms of the Serverless Trilemma, and the user experience was documented.

Nevertheless, this remains a proof of concept. But, even at this state, the resultant system already offers a lot of value to companies that rely heavily on serverless architecture. Workloads that only use standard libraries and don't rely heavily on file or networking are common in serverless.

*B. Future work*

As future work, the DMTCP implementation should be attempted with other cloud platforms' FaaS offerings. Also tests and adjustments should be made to guarantee compatibility with binaries that have third party dependencies. Benchmarking checkpoint time on different scenarios should be done too, such as when recycling the container or instance where the function was allocated initially, or with other storage technologies for the checkpoint files repository. Other checkpointing strategies may be evaluated too, like performing them at intervals instead of at the end. Lastly, duplicated executions should be completely prevented.